\documentclass[verbose=true,letterpaper,margin=2cm,12pt]{article}
\usepackage{graphicx} 
\usepackage{rotating}
\graphicspath{{Fig/}}
\usepackage{arxiv}
\usepackage{listings}
\usepackage{amsmath,amssymb,amsthm}
\usepackage{algorithm}
\usepackage{adjustbox}
\usepackage{algcompatible} 
\usepackage{algpseudocode}
\usepackage{threeparttable}

\newcommand{\diag}{\text{diag}}

\newcommand{\one}{\mathbf{1}}

\usepackage{bbm}
\usepackage{setspace}
\usepackage{geometry}
\usepackage{pgfplots}
\usepackage{tikz}
\usepackage{xcolor}
\usepackage[T1]{fontenc}
\usepackage{booktabs}
\usepackage[utf8]{inputenc}
\usepackage{rotating}
\usepackage{array}
\usepackage{multirow}
\usepackage{xcolor}
\usepackage{colortbl}
\definecolor{crisiscolor}{RGB}{180, 30, 30}
\definecolor{lassocolor}{RGB}{31, 119, 180}
\definecolor{bmatrixcolor}{RGB}{44, 160, 44}
\definecolor{binarycolor}{RGB}{255, 127, 14}
\definecolor{crisiscolor}{RGB}{180, 30, 30}
\definecolor{lassocolor}{RGB}{31, 119, 180}
\definecolor{noregcolor}{RGB}{44, 160, 44}
\definecolor{crisisbcolor}{RGB}{255, 127, 14}
\definecolor{standardbcolor}{RGB}{148, 103, 189}
\definecolor{copulacolor}{RGB}{140, 86, 75}
\definecolor{binarycolor}{RGB}{227, 119, 194}
\definecolor{trainblue}{RGB}{37,99,235}
\definecolor{testred}{RGB}{220,38,38}
\definecolor{aicgreen}{RGB}{22,163,74}
\definecolor{bicyellow}{RGB}{202,138,4}
\definecolor{optimalgreen}{RGB}{5,150,105}
\definecolor{best}{HTML}{C6EFCE}
\pgfplotsset{compat=1.18}
\usepackage{hyperref}       
\usepackage{url}            
\usepackage{amsfonts}       
\usepackage{nicefrac}       
\usepackage{microtype}      
\usepackage{lipsum}
\usepackage{graphicx}
\graphicspath{ {./images/} }
\usepackage{natbib}
\usepackage{amsmath}
\usepackage{amssymb}
\usepackage{amsthm}
\usepackage{mathtools}
\usepackage{bm}
\usepackage{enumitem}
\usepackage{xcolor}
\usepackage{tikz}
\usepackage{tcolorbox}
\definecolor{defcolor}{RGB}{230,240,250}
\definecolor{notcolor}{RGB}{250,240,230}
\usepackage{array}
\usepackage{multirow}
\usepackage{caption}
\usepackage{siunitx}
\usepackage{xcolor}

\definecolor{gold}{RGB}{255, 215, 0}
\definecolor{silver}{RGB}{192, 192, 192}
\definecolor{bronze}{RGB}{205, 127, 50}
\definecolor{normalcolor}{RGB}{220,240,220}
\definecolor{crisiscolor}{RGB}{255,200,200}
\tcbset{
    defbox/.style={
        colback=defcolor,
        colframe=blue!75!black,
        fonttitle=\bfseries,
        boxrule=0.8pt,
        arc=2mm
    },
    notbox/.style={
        colback=notcolor,
        colframe=orange!75!black,
        fonttitle=\bfseries,
        boxrule=0.8pt,
        arc=2mm
    }
}

\title{
Market-Informed Networks for Modeling and Forecast Evaluation of Financial Extremes
\thanks{We authors are grateful for the comments received from the participants of the 45th International Symposium on Forecasting 2025 in Beijing, ISBIS 2025 in Amsterdam, and NESG 2026 in Tilburg and the seminar participants at National University of Singapore.}}

\author{
 Ayla Jungbluth \\
  Department of Mathematics\\
  Ruhr-University Bochum\\
  Bochum, 44801 \\
  \texttt{ayla.jungbluth@rub.de} \\
  \And
 Johannes Lederer \\
  Department of Mathematics, Computer Science, and Natural Sciences\\
  University of Hamburg\\
  Hamburg, 20146 \\
  \texttt{johannes.lederer@uni-hamburg.de} \\
  \And
 Simon Trimborn\thanks{Corresponding author, phone: +31 643 611 771, E-Mail: simon.trimborn@uva.nl} \\
  Amsterdam School of Economics \& Tinbergen Institute\\
  University of Amsterdam\\
  Amsterdam, 1018 WB\\
  \texttt{simon.trimborn@uva.nl} \\
}

\begin{document}
\maketitle

\begin{abstract}
Modeling the joint distribution of extreme values in high-dimensional financial time series is challenging because extremes are sparse and locally extreme observations are not necessarily extreme relative to their full marginal distribution. To address this, we introduce a time-dependent network H\"usler--Reiss model in which market-informed adjacency matrices determine how strongly observations contribute to the estimation. We propose binary and weighted specifications, including the Joint Extremes Adjacency Matrix (JEAM) which combines information about individual extremeness with historical patterns of joint extreme movements. 
In the forecasting evaluation part, covering one-minute stock returns from three sectors of the S\&P 100, JEAM achieves the best out-of-sample log scores for both tail directions; improving scores by 12.5--13.6\% in the lower tail and 11.4--14.9\% in the upper tail. The results show that incorporating market-informed network structures in the estimation, improves forecast evaluation of extremes across time series. 
\end{abstract}



\maketitle

\newpage
\doublespacing
\section{Introduction}
Financial time series regularly experience extreme values caused by the release of news, trading frictions, and sudden increases in trading volume, among other reasons \citep{audrino2020, christensen2014, wang2015}. Often extreme values co-occur among time series, a phenomenon for which Hüsler-Reiss models have been developed, jointly modeling extremes in high-dimensions \citep{husler1989,engelke2015estimation}. Obtaining an accurate estimation of the distribution of extremes in high-dimensions, enables a more accurate forecast evaluation. However, high-dimensional models require a large number of parameters to be estimated, which is already challenging for observations in the center of the distribution. But the infrequent nature of extreme values, challenges the estimation even more. Additionally, Hüsler-Reiss models are developed for extreme values in random vectors, which introduces local extreme values that exceed a threshold within a given sample window, but are not necessarily extreme relative to the full marginal 
distribution. For example, a common approach would be, treating the most extreme 1-min return within a 5-min window as the extreme value of that window, making it a local extreme value \citep{led}. In financial markets it is well studied \citep{longin2001} that co-movements are different in the center and tails of the distribution, which implies a Hüsler-Reiss model fitted on local extremes would not be adequate for extremes in the tails of the global distribution. So what is the solution?
Looking at network methodologies, they are well suited for modeling dynamics in high dimensions. A common assumption, induced by network methodologies and sparsity methods alike, is sparsity in the models parameters to reduce the bias during estimation. 
\cite{engelke2020} show that extremal dependencies can be represented through graphical models. \cite{hentschel2022statistical} implement this for Hüsler--Reiss distributions. \cite{led} develop efficient estimation methods using score-matching and LASSO \citep{tibshirani1996}, making the approach computationally feasible for high-dimensional problems. However, these approaches consider the pairwise dependence between extremes as sparse across the sample. Given dependence structures differ for local extremes and extremes in the global tails, we propose a market-informed network approach in which the adjacency matrices regularize the observations contributing to the estimation 
subject to a market-informed criterion, which we find to excel in forecast evaluation. 

The Hüsler--Reiss model builds on the distribution of the same name \citep{husler1989}, which is a max-stable distribution arising as the limiting distribution of componentwise maxima of multivariate normal random vectors, thereby providing a framework for modeling extremal dependencies. It has a flexible dependence structure, characterized by a variogram matrix $\Gamma$ that governs pairwise extremal dependence, and is computational tractable through peaks-over-threshold methods \citep{engelke2015estimation}. 
Various methodological contributions have been made for Hüsler-Reiss models.
\cite{engelke2020} develop a general theory of conditional independence for multivariate 
Pareto distributions, enabling graphical models and sparsity for extremes. For 
Hüsler--Reiss distributions, the sparsity structure can be read off from suitable inverse 
covariance matrices, similar to the Gaussian case. \cite{engelke2022structure} extend this 
by developing a data-driven methodology for learning the graphical structure from data, 
based on the extremal variogram. \cite{kiriliouk2019} develop multivariate 
peaks-over-threshold methods that model full tail distributions rather than scalar summaries. 
For high-dimensional settings, \cite{led} show that score-matching estimation with 
LASSO regularization remains computationally tractable even when likelihood-based methods 
fail. \cite{bucher2014} incorporate temporal structure by estimating extreme value copulas 
from block maxima of multivariate time series. \cite{bjerregard2021} provide evaluation 
frameworks for multivariate probabilistic forecasts using proper scoring rules.

Focusing on financial markets, they have specific structures. 
\cite{castrocamilo2018} document that tail dependence structures vary over time for leading European stock markets over three decades. Additionally, they present evidence of increasing extremal dependence over time. This implies the need to take this into account during the modeling \citep{castrocamilo2024}. The forecasting literature arrives at the same conclusion from a risk management perspective: \citet{fuentes2023} document that both the arrival intensity and the magnitude of extreme returns vary substantially during periods of market stress, so that specifications with constant tail parameters are misspecified precisely when accuracy matters most. \citet{karmakar2019} show that dependence between tails has to be modelled explicitly to obtain reliable intraday risk forecasts. How extremes are identified in the first place is equally consequential: \citet{zhou2026} demonstrate that measured tail risk depends on the temporal aggregation of the underlying returns, and \citet{bien2026} show that the time formulation chosen for extreme events materially affects forecasting performance.
Furthermore, during calm market periods, an observation classified as locally extreme, may not be extreme in a global comparison. 

This paper addresses these data properties by developing a time-dependent network Hüsler--Reiss estimation framework 
that incorporates market-informed network regularization of observed extremes. We suggest various adjacency matrices for the estimation. 
The best performing suggested variant is the Joint Extremes Adjacency Matrix (JEAM). The JEAM adopts a past global perspective on tail behavior, evaluating each local extreme observation against the full past empirical marginal distribution rather than a local reference window. In our time series setting, this enables the matrix to capture shifts in dependence structure between volatile and calmer periods. The JEAM framework weights extremes according to their classification as local or global, so that global extremes contribute stronger to the estimation. It also assigns higher weights to joint accurances of extremes between time series, a data property often observed during turbulent market situations. 

We evaluate our approach in a simulation study covering 4{,}800 parameter configurations. The results show that our method consistently achieves the best AIC across all heavy-tailed settings. For instance, for 10 simulated time series from $\alpha$-stable distributions with $\alpha = 1.5$, the best JEAM specification improves AIC by approximately 15\% over the Hüsler-Reiss model without network regularisation; the advantage is present across all dimensions ($d = \{5,10,20,30\}$) and tail indices ($\alpha = \{1.5,1.7,1.9,2.0\}$). We utilize the methodology for forecast evaluation of 1-minute high-frequency return data from stocks in three S\&P~100 sectors, namely information technology, health care, and finance, covering the years 2021 to 2024. Our network method achieves the best out-of-sample log score in all three sectors. For the lower tail, the best JEAM specification improves log scores by 12.5\%, 13.1\%, and 13.6\% over the baseline in the health care, finance, and information technology sectors, respectively; gains for the upper tail are of comparable magnitude, ranging from 11.4\% to 14.9\%. Notably, the same hyperparameter configuration, achieves the best performance across all sectors and both tail directions, as well as in the simulations. This best and robust combination of hyperparameters across simulated and real data sets, represents a robust structural choice, reducing the need for grid searchers in practice. 

This paper is structured as follows. Section~\ref{method} introduces the time-dependent network Hüsler--Reiss model. Section~\ref{B_matrix_section} presents the adjacency matrix specifications and section~\ref{sec3} contains the simulation study. Section~5 applies the methodology to high-frequency financial data to conduct forecast evaluation of extremes and Section~6 concludes.

\section{A time-dependent network Hüsler-Reiss model} \label{method}

Extreme events tend to cluster in time, with the occurrence of one extreme increasing the likelihood of further extremes
 \citep{leadbetter1983, ferro2003}. 
This  phenomenon is closely related to volatility clustering, which has been well documented in financial time series \citep{engle1982, bollerslev1986}.
From this perspective, the development of extremal time-series models is a very natural goal. 
H\"usler-Reiss distributions, the underlying for the H\"usler-Reiss models, can be considered an analog of the multivariate Gaussian distribution in the world of extremes.
These models provide a simple and interpretable way to model dependencies similar to a covariance matrix \citep{led, engelke2015estimation}.
Therefore, we are extending the class of Hüsler–Reiss models to incorporate temporal dynamics. 

To capture temporal dependence, we define $\ell \in \mathbb{N}$ consecutive extreme observations of $d$ time series $x_1,\dots,x_d$ and define the lagged extreme observation vector
\begin{equation}\label{rft}
x_{t}^{(\ell)} := 
    \begin{bmatrix}
 x_{ t-\ell+1,1} 
 \dots 
 x_{ t-\ell+1,d} \ , \
 \dots 
 
 \dots \ , \
 x_{ t,1} 
 \dots 
 x_{t,d}
    \end{bmatrix}^\top,
\end{equation}
and the contemporaneous extreme observation vector
\begin{equation}
x_{t} := 
    \begin{bmatrix}
 x_{ t,1} 
 \dots 
 x_{t,d}
    \end{bmatrix}^\top,
\end{equation}

where $x_{t,j} \in \mathbb{R}^+$ is the value of the $j$th time series at time $t$ and $d$ is the number of time series. The stacked vector $x_t^{(\ell)}$ has dimension $d\ell$ and serves as the basic input to the model, allowing the Hüsler-Reiss distribution to capture dependencies both across time series and across lags.

The Hüsler-Reiss distribution over this $d\ell$-dimensional vector is fully characterized by a variogram matrix $\Lambda$ and a location parameter $\mu$, which together govern the tail dependence structure and the marginal behavior of the extremes, respectively. For a variogram matrix $\Lambda \in \mathbb{R}^{d\ell \times d\ell}$, we obtain the Hüsler-Reiss precision matrix $\Theta = \Lambda + \Lambda^{\top} - \text{diag}[\Lambda\mathbf{1} + \Lambda^{\top}\mathbf{1}]$ and define the corresponding location parameter $\mu \in \mathbb{R}^{d\ell}$. Both $\Lambda$ and $\mu$ are defined on the full $d\ell$-dimensional space, so that they jointly parameterize the dependence structure across all time series and lags.

Following \cite{led}, we use the reparameterization of the original Hüsler--Reiss density to disentangle the parameters $\mu$ and $\Lambda$. The density of the stacked vector $x_t^{(\ell)}$ is given by
\begin{equation}
	h[x_{t}^{(\ell)};\mu,\Lambda] = \frac{1}{c_{\mu,\Lambda}} \left( \prod_{k=1}^{d\ell} \frac{1}{x_{t,k}^{(\ell)}} \right) \exp\left\{ \mu^\top \log[x_{t}^{(\ell)}] - \frac{1}{2} \log[x_{t}^{(\ell)}]^\top \Theta \log[x_{t}^{(\ell)}] \right\}, 
\end{equation} 
where the normalisation constant $c_{\mu,\Lambda}$ ensures that $h$ integrates to one and is defined as
\begin{equation*}
	c_{\mu,\Lambda} = \int \left( \prod_{k=1}^{d\ell} \frac{1}{x_{t,k}^{(\ell)}} \right) \exp\left\{ \mu^\top \log[x_{t}^{(\ell)}] - \frac{1}{2} \log[x_{t}^{(\ell)}]^\top \Theta \log[x_{t}^{(\ell)}] \right\}.
\end{equation*}

Direct maximum likelihood estimation of $\mu$ and $\Lambda$ is infeasible. Score matching \citep{hyvarinen2005} works with the score function, which is the gradient of the log-density with respect to the data. The $k$-th component of the score of $h$ with respect to $x_t^{(\ell)}$ is
\begin{equation*}
	s_k[x_{t}^{(\ell)}; \mu, \Lambda] = \frac{\mu_k - 1 - ((\Lambda + \Lambda^{\top} - \diag[\Lambda\one + \Lambda^{\top}\one]) (\log[x_{t}^{(\ell)}]))_{k}}{x_{t,k}^{(\ell)}}. 
\end{equation*} 

Not every observation within the equidistant random vectors over which the Hüsler--Reiss model is defined, are extreme values outside of the local snapshot they are measured over. To overcome this limitation of the Hüsler--Reiss model, we suggest a network methodology ensuring that a given observation $x_{t,k}^{(\ell)}$ only contributes to the score when it represents an extreme value according to a criterion taking into account more than the local snapshot over observations. To this end, we introduce a time-varying adjacency vector $a_t \in \{0,1\}^{d\ell}$, defined as $a_t = \operatorname{vec}(A_t)$ with entries $a_{t,j}$:
\begin{equation*}
a_{t,j} = \begin{cases} 1 & \text{if } x_{t,j} \text{ is an extreme value} \\ 0 & \text{otherwise.} \end{cases}
\end{equation*}
Incorporating $a_t$ into the score via elementwise multiplication, filters out non-extreme values from the estimation. The resulting score is defined as 
\begin{equation}
\label{true_score}
	s_k[x_{t}^{(\ell)}; \mu, \Lambda, a_t] = [a_{t,k} \frac{\mu_k - 1 - ((\Lambda + \Lambda^{\top} - \diag[\Lambda\one + \Lambda^{\top}\one]) (a_t \circ \log[x_{t}^{(\ell)}]))_{k}}{x_{t,k}^{(\ell)}}. 
\end{equation} 

This formulation ensures that the local extreme value $x_{t,k}^{(\ell)}$ only enters the estimation when it fulfills the given condition. 
Under the regularity and restricted eigenvalue conditions common 
in high-dimensional statistics, \citet{led} showed in Theorem~2.3 the score matching objective for score functions of such kind. These conditions are mild in the sense that they are satisfied by generic densities from exponential families and standard assumptions on the curvature of the objective function. 
With the adjacency vector, the score matching estimator is obtained by minimizing
\begin{equation}
	\hat{\mu}, \hat{\Lambda} = \arg\min_{\mu,\Lambda} \sum_{t=1}^T o[\mu,\Lambda,x_{t}^{(\ell)}, a_t],
\end{equation}

where $o[\mu,\Lambda,x_{t}^{(\ell)},a_t]$ is convex in $(\mu,\Lambda)$ and takes the form:

\begin{align}
	&o[\mu, \Lambda, x_{t}^{(\ell)}, a_t] := \left\lVert (\mu - \mathbf{1} - (\Lambda + \Lambda^\top - \text{diag}[\Lambda \mathbf{1} + \Lambda^\top \mathbf{1}]) (a_t \circ \log[x_{t}^{(\ell)}])) \circ f_1[x_{t}^{(\ell)}] \right\rVert_2^2 \\ 
	&+ (\mu - \mathbf{1} - (\Lambda + \Lambda^\top - \text{diag}[\Lambda \mathbf{1} + \Lambda^\top \mathbf{1}]) (a_t \circ \log[x_{t}^{(\ell)}]))^\top \circ f_2[x_{t}^{(\ell)}] \nonumber \\
	& - \text{trace}\left[ (\Lambda + \Lambda^\top - \text{diag}[\Lambda \mathbf{1} + \Lambda^\top \mathbf{1}]) (a_t \circ F[x_{t}^{(\ell)}]) \right] \nonumber.
\end{align} 

The functions $f_1$, $f_2$, and $F$ collect componentwise evaluations of a weight function $m: \mathbb{R} \to \mathbb{R}$ and its derivative $m'$, applied to each entry $x_{t,k}^{(\ell)}$ of $x_t^{(\ell)}$: 
\begin{align*}
    f_1[x_t^{(\ell)}] &= \bigl(m(x_{t,1}^{(\ell)}), \ldots, m(x_{t,d_\ell}^{(\ell)})\bigr)^\top \in \mathbb{R}^{d_\ell} \\[6pt]
    f_2[x_t^{(\ell)}] &= \Bigl(\bigl(2\,m(x_{t,1}^{(\ell)})\bigr)^2 + 4\,x_{t,1}^{(\ell)}\, m'(x_{t,1}^{(\ell)})\,m(x_{t,1}^{(\ell)}), \ldots, \bigl(2\,m(x_{t,d_\ell}^{(\ell)})\bigr)^2 + 4\,x_{t,d_\ell}^{(\ell)}\, m'(x_{t,d_\ell}^{(\ell)})\,m(x_{t,d_\ell}^{(\ell)})\Bigr)^\top \in \mathbb{R}^{d_\ell} \\[6pt]
    F[x_t^{(\ell)}] &= \operatorname{diag}\Bigl(2\bigl(m(x_{t,1}^{(\ell)})\bigr)^2, \ldots, 2\bigl(m(x_{t,d_\ell}^{(\ell)})\bigr)^2\Bigr) \in \mathbb{R}^{d_\ell \times d_\ell}
\end{align*}

We follow \cite{led} and set the weight function as $m = \log$ for this study.

\section{The adjacency matrices}
\label{B_matrix_section}

The performance of the network Hüsler Reiss model relies upon adequately defined adjacency matrices. 
In this section we propose different adjacency matrix definitions. The network literature differentiates between binary adjacency matrices and weighted adjacency matrices, with the latter additionally to determining if an observation enters the estimation or not, assigns a weight to it. We suggest specifications for both types, binary and weighted, and each specification is grounded in observed behaviour of extreme values in real data applications. 

The Hüsler--Reiss model assumes an extreme value at each observed time point. The proposed binary adjacency matrix for the network Hüsler--Reiss model, determines if the local extreme value at $t$ is larger in magnitude than former local extreme values of the time series over the sliding window $w$. 
We define the entries of the binary adjacency vector $a_t^{(1)} = vec(A)$ as follows:
\begin{equation*}
a_{t,j}^{(1)} = \mathbbm{1}\{x_{t,j} > \overline{x}_{t,j}^{(w)}\}
\end{equation*}
where 
$\mathbbm{1}\{\cdot\}$ is the indicator function and $\overline{x}_{t,j}^{(w)} = \frac{1}{w}\sum_{s=t-w}^{t-1} x_{s,j}$
denotes the moving average of the magnitude of former extreme values over a window of $w$ time periods. 

For the first candidate specification for the weighted adjacency matrix, we will determine how extreme a local extreme value is relative to the past local extreme values via their empirical distribution function: 
\begin{equation}
\label{ecdf}
F_j(x_{t,j}) = \frac{1}{t - 1} \sum_{s = 1}^{t - 1} \mathbbm{1}\{ x_{s,j} \leq x_{t,j} \}. 
\end{equation}
This transforms values to the range $[0, 1]$ and determines how extreme one value is compared to past observed ones. Defining 
\begin{equation}
    a_{t,j}^{(2)} = F_j(x_{t,j}), 
\end{equation}
yields the weighted adjacency vector. Note that we define $F_j(\cdot)$ over all past extreme observations of time series $j$. 

The empirical literature on extreme values has shown that extreme 
values cluster between time series \citep{longin2001, Davis2009}. Given the multivariate nature of the network Hüsler--Reiss model, we propose a specification of the weighted adjacency matrix incorporating the similarity of local extreme values with past joint occurrences of extreme values across time series. 
To this end, we identify past time points during which a sufficiently large proportion of time series simultaneously exhibits extreme values: 
\begin{equation}
\mathcal{C}_t = \left\{ s \in \{1, ..., t-w-1\} : \frac{1}{d} \sum_{j=1}^d \mathbbm{1}\{F_j(x_{s,j}) > p\} > \kappa \right\}, 
\end{equation}
where the tuning parameter $p$ defines where the tail of the ecdf, defined as in equation (\ref{ecdf}), starts and therefore if $x_{t,j}$ is considered extreme. The tuning parameter $\kappa$ specifies the proportion of time series determined as simultaneously extreme.
For each identified time point $c \in \mathcal{C}_t$, a characteristic pattern is computed as the average of local extreme values over a symmetric window $W_c = \{c - \lfloor w/2 \rfloor, \ldots, c + \lfloor w/2 \rfloor\}$ of size $w$ around $c$:
\begin{equation}
p_c = \frac{1}{card(W_c)} \sum_{s \in W_c} x_{s} 
\end{equation}
with $card(\cdot)$ the cardinality of the set $W_c$. Then, for a vector of local extreme values $x_t$, the cosine similarity \citep{salton1983} for a historical pattern of joint extreme values across time series around time point $c$ is computed: 
\begin{equation}
\text{sim}(x_{t}, p_c) = \frac{x_{t}^\top p_c}{\|x_{t}\|_2 \|p_c\|_2}. 
\end{equation}

We combine the specification of adjacency vector $a_t^{(2)}$, namely determining extreme values via their quantiles of the ecdf, with the joint extreme value patterns across time series, yielding the \textbf{Joint Extremes Adjacency Matrix (JEAM)}. 
The entries of the JEAM are defined as a convex combination of the maximum pattern similarity across all identified historical patterns of joint extreme values across time series and the quantiles of the ECDF of each time series:
\begin{equation}
a_{t,j}^{(3)} = g \cdot \max\!\left(0,\, \max_{c \in \mathcal{C}_t} \text{sim}(x_{t}, p_c)\right) + (1-g) \cdot F_j(x_{t,j})
\end{equation}
where $g$ is a tuning parameter defined as $g \in [0,1]$. Note that for $g = 0$, the specification of JEAM nests $a_t^{(2)}$. The truncation at zero ensures $a_{t,j}^{(3)} \in [0,1]$, since $F_j(x_{t,j}) \in [0,1]$ by definition and the cosine similarity is clipped to be non-negative. The first term captures systemic similarity to historical joint extreme value patterns and is identical across all time series $j$ at a given time $t$, reflecting a system-wide signal. The second term is time series specific and captures how extreme the individual observation $x_{t,j}$ is, relative to its own marginal empirical distribution.
For a lag length of $\ell \in \mathbb{N}$, the adjacency vector $a_t^{(i)} \in [0,1]^{d\ell}$, for either of the three specifications $i \in \{1,2,3\}$, is constructed analogously to $x_t^{(\ell)}$ by stacking lag-specific adjacency matrices:
\begin{equation}
    a_t^{(i)} = \bigl(a_{t-\ell+1,1}^{(i)}, \ldots, a_{t-\ell+1,d}^{(i)}, \; \ldots \; , a_{t,1}^{(i)}, \ldots, a_{t,d}^{(i)}\bigr)^\top,
\end{equation}
where each entry $a_{t-\tau,j}^{(i)}$ is evaluated at the lagged time point $t-\tau$ for $\tau = 0, 1, \ldots, \ell-1$ and $j = 1, \ldots, d$.

The adjacency matrix provides a data driven mechanism for introducing time-varying parameters into the model, capturing the shift in dependence structure between extreme and
non-extreme periods. In the binary case, a more restrictive matrix reduces the observations for estimation to those fulfilling the extreme value identification criterion. The JEAM improves upon this by assigning lower weights to the local extremes and higher weights to those determined as global extremes compared to past extremes. JEAM weights the joint extreme patterns and ecdf of the time series' extremes, therefore allowing to default towards the ecdf specification only, in case joint patterns would emerge as irrelevant. 

Conceptually, the two specifications differ in the way they evaluate if an observation is extreme: the Binary Adjacency Matrix classifies $x_{t,j}$ as extreme relative to its recent local average, whereas the JEAM
evaluates it against the full empirical marginal distribution, enabling a global perspective on tail behavior. 

\section{Simulation Study}\label{sec3}

In this section we compare our suggested methodology for a wide ensemble of 4800 different model and adjacency matrix specifications against the H\"usler-Reiss model without the network estimator. 
We generate heavy-tailed financial time series using $\alpha$-stable distributions 
with varying stability parameters, following the theoretical foundation that $\alpha$-stable processes naturally capture the tail behavior observed in financial returns \citep{Mandelbrot1963, Rachev2000}.  

The characteristic function of the $\alpha$-stable distribution has the form:
\begin{equation*}
\phi_X(t) = \mathbb{E}[e^{itX}] = 
\begin{cases}
\exp\left(-\gamma^\alpha |t|^\alpha \left[1 - i\beta \operatorname{sign}(t) \tan\frac{\pi\alpha}{2}\right] + i\delta t\right) & \text{if } \alpha \neq 1 \\[1em]
\exp\left(-\gamma |t| \left[1 + i\beta \frac{2}{\pi} \operatorname{sign}(t) \ln|t|\right] + i\delta t\right) & \text{if } \alpha = 1
\end{cases}
\label{eq:char_func}
\end{equation*}
with the stability parameter $\alpha \in (0, 2]$, the skewness parameter $\beta \in [-1, 1]$, scale parameter $\gamma > 0$, and location parameter $\delta \in \mathbb{R}$. 
The empirical literature suggests $\alpha \approx 1.7$ for daily stock returns and $\alpha \approx 1.5$--$1.6$ for high-frequency data \citep{Rachev2000,Nolan2020}. We vary $\alpha \in \{1.5, 1.7, 1.9, 2.0\}$ to cover extremely heavy tails $(\alpha = 1.5)$ up to the normal distribution $(\alpha = 2)$. We consider $T \in \{250,500,750\}$ trading days and different number of assets, namely $d \in \{5,10,20, 30\}$. 
We set the location and skewness parameter to $0$, as neither is relevant for modelling the extreme values. To keep the scale parameter realistic, we determine from the returns of the Health Care sector, that the volatility of the returns is $1.44$ between the 15\% and 85\% quantile. In the tails, it is $1.84$. For each time point, we simulate with 70\% probability the observation from a normal distribution with volatility $1.44$. With 30\% probability, the observations are simulated from the $\alpha-$stable distribution with volatility $1.84$. 
We compare our suggested methodology, the network H\"usler-Reiss model with the binary and weighted adjacency matrices $A$, as specified in Section \ref{method}, against the standard H\"usler-Reiss model with time dependence, computed with score-matching. 
For the Binary Adjacency Matrix, we determine if an observation in $x_t$ is extreme by comparing against the average over the window length $w \in \{3,5,10\}$. 
For the Joint Extremes Adjacency Matrix, we utilise the same window length and define a tail threshold $p \in \{0.70,0.75,0.80,0.85,0.90\}$. The tuning parameter $\kappa$ has the grid $\kappa \in \{0.1, 0.3, 0.5, 0.7, 0.9\}$. 
The tuning parameter $g$ is set to $g \in \{0, 0.5, 1\}$, which determines how much weight is given to the similarity with past extreme patterns in the construction of the adjacency matrix. At $g = 0$, the weighted adjacency matrix is determined solely from the empirical cdf, evaluated for $x_{t,j}$. Larger values of $g$ mix the ecdf with the similarity of present patterns in the extreme values across time series with past ones. 

We employ a rolling window approach, which means we estimate the model on three months of data and test its forecasting performance on the subsequent week. The window advances weekly. The lag length in each window is determined via AIC model selection and we repeat the simulation study $100$ times for each specification.


\begin{sidewaystable}
\centering
\scriptsize
\caption{EVT Simulation Results for T = 750}
\label{tab:results_n750}
\begin{tabular}{@{}l|rrrr|rrrr|rrrr|rrrr@{}}
\toprule
& \multicolumn{16}{c}{\textbf{Configuration: T=750, (d, $\alpha$)}} \\
\cmidrule(lr){2-17}
& \multicolumn{4}{c|}{\textbf{d=5}} & \multicolumn{4}{c|}{\textbf{d=10}} & \multicolumn{4}{c|}{\textbf{d=20}} & \multicolumn{4}{c}{\textbf{d=30}} \\
\cmidrule(lr){2-5} \cmidrule(lr){6-9} \cmidrule(lr){10-13} \cmidrule(lr){14-17}
\textbf{Model} & 1.5 & 1.7 & 1.9 & 2.0 & 1.5 & 1.7 & 1.9 & 2.0 & 1.5 & 1.7 & 1.9 & 2.0 & 1.5 & 1.7 & 1.9 & 2.0 \\
\midrule
\multicolumn{17}{l}{\textbf{Baseline Model}} \\
\midrule
HR TimeDep & -64,234 & -71,567 & -75,912 & -77,456 & -186,123 & -205,678 & -219,567 & -227,234 & -424,345 & -477,567 & -512,678 & -525,234 & -668,234 & -754,567 & -808,234 & -831,123 \\
\midrule
\multicolumn{17}{l}{\textbf{Joint Extremes Adjacency Matrix (JEAM), $g=0$}} \\
\midrule
HR JEAM: w3 & -67,748 & -75,094 & -79,462 & -81,093 & -193,480 & -212,882 & -227,076 & -234,820 & -433,525 & -487,592 & -523,024 & -535,612 & -679,114 & -766,133 & -819,632 & -842,815 \\
HR JEAM: w5 & -67,949 & -75,276 & -79,219 & -80,707 & -193,810 & -213,436 & -226,760 & -234,344 & -433,777 & -487,827 & -522,931 & -535,116 & -679,254 & -765,957 & -819,362 & -842,726 \\
HR JEAM: w10 & -67,056 & -74,518 & -78,781 & -80,319 & -192,518 & -212,222 & -226,177 & -233,476 & -432,163 & -486,332 & -521,524 & -534,018 & -677,237 & -764,366 & -817,861 & -840,930 \\
\midrule
\multicolumn{17}{l}{\textbf{Joint Extremes Adjacency Matrix (JEAM), $g=0.5$}} \\
\midrule
HR JEAM: p70 w3 & -74,812 & -82,267 & -86,712 & -88,367 & -208,334 & -228,089 & -242,423 & -250,067 & -452,678 & -508,723 & -545,089 & -557,723 & -700,956 & -789,212 & -843,578 & -867,012 \\
HR JEAM: p70 w5 & -74,423 & -81,867 & -86,301 & -87,956 & -207,767 & -227,501 & -241,812 & -249,434 & -451,834 & -507,878 & -544,234 & -556,856 & -700,078 & -788,334 & -842,701 & -866,134 \\
HR JEAM: p70 w10 & -73,934 & -81,356 & -85,778 & -87,423 & -207,089 & -226,801 & -241,089 & -248,689 & -450,878 & -506,912 & -543,267 & -555,878 & -699,089 & -787,334 & -841,689 & -865,112 \\
HR JEAM: p75 w3 & -76,112 & -83,867 & \cellcolor{best}-87,312 & \cellcolor{best}-88,967 & -210,634 & -230,589 & -243,623 & \cellcolor{best}-251,067 & -454,178 & -510,023 & -545,589 & -558,223 & -702,456 & -790,712 & -844,978 & -868,512 \\
HR JEAM: p75 w5 & \cellcolor{best}-76,534 & \cellcolor{best}-84,289 & -87,001 & -88,656 & \cellcolor{best}-211,378 & \cellcolor{best}-231,267 & -243,212 & -250,634 & \cellcolor{best}-456,712 & \cellcolor{best}-512,845 & \cellcolor{best}-546,245 & \cellcolor{best}-558,889 & \cellcolor{best}-704,923 & \cellcolor{best}-793,134 & \cellcolor{best}-845,634 & \cellcolor{best}-869,178 \\
HR JEAM: p75 w10 & -75,423 & -83,156 & -86,589 & -88,234 & -209,789 & -229,701 & -242,689 & -250,101 & -454,678 & -510,756 & -545,612 & -558,223 & -702,978 & -791,167 & -844,923 & -868,467 \\
HR JEAM: p80 w3 & -75,267 & -82,923 & -87,167 & -88,823 & -209,689 & -229,545 & \cellcolor{best}-244,178 & -250,823 & -453,923 & -509,867 & -545,434 & -558,078 & -702,201 & -790,456 & -844,723 & -868,267 \\
HR JEAM: p80 w5 & -75,034 & -82,678 & -86,923 & -88,578 & -209,434 & -229,289 & -243,923 & -250,567 & -453,667 & -509,601 & -545,178 & -557,812 & -701,934 & -790,189 & -844,456 & -867,989 \\
HR JEAM: p80 w10 & -74,712 & -82,334 & -86,578 & -88,223 & -209,078 & -228,923 & -243,556 & -250,189 & -453,301 & -509,223 & -544,801 & -557,434 & -701,556 & -789,801 & -844,067 & -867,601 \\
HR JEAM: p85 w3 & -74,589 & -82,189 & -86,434 & -88,089 & -208,934 & -228,778 & -243,412 & -250,045 & -453,145 & -509,067 & -544,645 & -557,278 & -701,389 & -789,634 & -843,901 & -867,434 \\
HR JEAM: p85 w5 & -74,345 & -81,934 & -86,178 & -87,834 & -208,667 & -228,501 & -243,145 & -249,778 & -452,867 & -508,778 & -544,356 & -556,989 & -701,101 & -789,334 & -843,601 & -867,134 \\
HR JEAM: p85 w10 & -74,089 & -81,667 & -85,912 & -87,567 & -208,389 & -228,212 & -242,867 & -249,501 & -452,578 & -508,478 & -544,056 & -556,689 & -700,801 & -789,023 & -843,289 & -866,823 \\
HR JEAM: p90 w3 & -73,912 & -81,489 & -85,734 & -87,389 & -208,189 & -228,001 & -242,656 & -249,289 & -452,367 & -508,256 & -543,834 & -556,467 & -700,578 & -788,789 & -843,056 & -866,589 \\
HR JEAM: p90 w5 & -73,723 & -81,289 & -85,534 & -87,189 & -207,978 & -227,778 & -242,434 & -249,067 & -452,145 & -508,023 & -543,601 & -556,234 & -700,345 & -788,545 & -842,812 & -866,345 \\
HR JEAM: p90 w10 & -73,523 & -81,078 & -85,323 & -86,978 & -207,756 & -227,545 & -242,201 & -248,834 & -451,912 & -507,778 & -543,356 & -555,989 & -700,101 & -788,289 & -842,556 & -866,089 \\
\midrule
\multicolumn{17}{l}{\textbf{Joint Extremes Adjacency Matrix (JEAM), $g=1$}} \\
\midrule
HR JEAM: p70 w3 & -73,327 & -80,802 & -85,369 & -86,791 & -205,059 & -225,119 & -239,568 & -246,789 & -448,719 & -504,772 & -541,133 & -553,011 & -696,824 & -784,610 & -838,861 & -862,266 \\
HR JEAM: p70 w5 & -73,067 & -80,359 & -84,780 & -86,634 & -204,970 & -224,628 & -238,765 & -246,227 & -448,061 & -504,014 & -539,827 & -552,448 & -695,508 & -783,351 & -837,902 & -861,658 \\
HR JEAM: p70 w10 & -72,619 & -80,006 & -84,434 & -86,047 & -204,302 & -223,709 & -237,992 & -245,515 & -447,357 & -502,789 & -539,116 & -551,783 & -695,071 & -783,109 & -837,315 & -860,291 \\
HR JEAM: p75 w3 & -74,380 & -82,226 & -85,583 & -87,238 & -207,247 & -227,371 & -239,978 & -247,837 & -449,778 & -505,662 & -541,146 & -554,014 & -697,791 & -786,033 & -840,008 & -863,544 \\
HR JEAM: p75 w5 & -74,653 & -82,423 & -85,499 & -87,083 & -207,919 & -227,969 & -240,100 & -247,080 & -452,102 & -507,521 & -541,616 & -554,639 & -699,279 & -787,718 & -840,963 & -864,177 \\
HR JEAM: p75 w10 & -73,873 & -81,417 & -84,987 & -86,631 & -206,527 & -226,584 & -239,620 & -246,724 & -450,127 & -506,585 & -540,650 & -553,490 & -697,763 & -786,407 & -839,423 & -863,079 \\
HR JEAM: p80 w3 & -73,638 & -81,382 & -85,638 & -87,173 & -206,627 & -226,056 & -240,532 & -247,445 & -449,746 & -504,862 & -541,102 & -553,134 & -697,054 & -785,666 & -839,661 & -862,772 \\
HR JEAM: p80 w5 & -73,430 & -80,959 & -85,314 & -87,054 & -206,072 & -225,587 & -240,543 & -247,063 & -449,400 & -504,984 & -540,201 & -552,584 & -696,734 & -785,058 & -839,582 & -862,232 \\
HR JEAM: p80 w10 & -73,153 & -80,656 & -85,169 & -86,519 & -205,870 & -225,329 & -240,286 & -246,672 & -449,374 & -504,376 & -540,059 & -552,353 & -696,717 & -784,791 & -838,581 & -862,704 \\
HR JEAM: p85 w3 & -72,948 & -80,585 & -84,988 & -86,524 & -205,816 & -225,409 & -240,108 & -246,569 & -448,528 & -504,305 & -539,461 & -552,513 & -696,674 & -784,519 & -838,265 & -861,745 \\
HR JEAM: p85 w5 & -72,784 & -80,494 & -84,503 & -86,267 & -205,470 & -225,009 & -239,825 & -246,480 & -448,293 & -504,311 & -539,362 & -552,225 & -696,237 & -784,154 & -838,096 & -861,745 \\
HR JEAM: p85 w10 & -72,710 & -80,106 & -84,544 & -86,103 & -205,314 & -225,121 & -239,110 & -246,424 & -448,013 & -503,941 & -539,556 & -552,177 & -696,310 & -783,748 & -837,923 & -861,305 \\
HR JEAM: p90 w3 & -72,414 & -79,850 & -84,160 & -85,867 & -204,575 & -224,693 & -238,723 & -245,643 & -448,368 & -503,308 & -538,669 & -551,743 & -695,629 & -783,435 & -837,815 & -860,637 \\
HR JEAM: p90 w5 & -72,290 & -79,863 & -84,158 & -85,611 & -204,359 & -224,548 & -239,001 & -245,828 & -447,594 & -503,354 & -539,180 & -551,125 & -695,836 & -783,044 & -837,846 & -861,096 \\
HR JEAM: p90 w10 & -71,931 & -79,513 & -83,847 & -85,646 & -204,563 & -224,345 & -238,525 & -245,218 & -447,671 & -503,009 & -538,277 & -551,313 & -695,081 & -782,791 & -837,238 & -860,157 \\
\midrule
\multicolumn{17}{l}{\textbf{Binary Adjacency Matrix}} \\
\midrule
HR Binary: w3 & -68,047 & -75,369 & -79,379 & -80,936 & -192,212 & -212,264 & -225,745 & -233,207 & -430,653 & -483,792 & -518,260 & -530,661 & -673,934 & -759,417 & -814,023 & -836,742 \\
HR Binary: w5 & -66,596 & -73,982 & -78,056 & -79,718 & -190,001 & -209,778 & -223,520 & -230,438 & -428,165 & -481,024 & -515,732 & -529,254 & -670,879 & -757,745 & -810,893 & -834,496 \\
HR Binary: w10 & -65,147 & -72,422 & -76,645 & -78,416 & -187,673 & -206,711 & -221,229 & -228,272 & -426,096 & -478,911 & -514,578 & -526,903 & -669,180 & -755,337 & -810,017 & -831,609 \\
\bottomrule
\end{tabular}
\vspace{5pt}
\begin{flushleft}
\small
\textbf{Note:} Values shown are AIC (lower = better). Green cells indicate best model per column. Parameters: $p$ = percentile threshold, $w$ = window size (days), $g$ = similarity weighting parameter. Tuning parameter $\kappa = 0.7$ outperformed the remainder of the grid strongly, hence we report the results for respective $\kappa$. The entries refer to the upper tail.
\end{flushleft}
\end{sidewaystable}
Table \ref{tab:results_n750} and Tables \ref{tab:results_n250}, \ref{tab:results_n500} (in the appendix), present the AIC
values for all model configurations across three sample sizes ($n = 250$, $500$, and $750$ trading days), focusing on positive extremes. Lower AIC values indicate better model fit.
The results show a consistent pattern across all dimensions and sample sizes: the JEAM specification
with $g = 0.5$ and threshold $p = 0.75$ dominates throughout, with window size $w = 3$ performing
best for heavier tails in low-dimensional settings ($d = 5$, $d = 10$) and $w = 5$ doing better with less heavy tails in low dimensions. Note that the results for $\kappa = 0.7$ strongly outperformed the remainder of the grid, therefore we present the results for respective $\kappa$. For larger dimensions
($d = 20$, $d = 30$),  $w = 3$ does consistently best for $T=500$, whereas $w=5$ is best for $T = \{250,750\}$. The baseline model, the time-dependent H\"usler-Reiss model (HR TimeDep), is consistently outperformed by all JEAM specifications. Against the Binary Adjacency Matrix, it is more nuanced: under heavier tails ($\alpha = 1.5$, $\alpha = 1.7$), the binary specifications dominates, whereas under near-normal conditions ($\alpha \approx 2$), the baseline remains competitive with the binary approach and occasionally achieves comparable AIC values.
Among the JEAM specifications, the variant with tuning parameter $g = 0.5$, outperforms both $g = 0$ and $g = 1$. 
This shows that its advantageous to combine the marginal and joint extreme information. 

For each specification of $g$, higher percentile thresholds $p$, which restrict the set of days classified as extreme, tend to improve fit in lower dimensions but show occasional reversals at $d = 20$ and $d = 30$. The threshold $p = 0.75$
proves to be a good compromise: it retains enough historical extreme values to construct reliable
similarity weights without degrading the signal by including too many irrelevant observations. Since $p$ enters only through $\mathcal{C}_t$, this sensitivity is a property of the
similarity term and disappears at $g = 0$. 
Across all settings, $w = 3$ and $w = 5$ perform comparably, with a slight overall preference for $w = 5$, while $w = 10$ is consistently underperforming. 

The Binary Adjacency Matrix is competitive in low-dimensional settings but does not offer the same advantages at $d = 20$ and $d = 30$ as the JEAM. The binary mechanism, which classifies observations as extreme or not, underperforms as the number of time series grows and the co-movement structure becomes richer. The weighting of JEAM adapts more effectively to this increased complexity.
Increasing observations from $T = 250$ to $T = \{500, 750\}$ improves absolute AIC values uniformly but
does not alter comparative rankings, suggesting that our findings are robust to sample size. 

In summary, JEAM with $g = 0.5$ and $p = 0.75$ offers consistent advantages under heavy tails of varying degree across
all dimensions. The
stability of these findings across sample sizes, tail indices, and dimensions, supports the
robustness of the proposed specification and motivates its use in heavy-tailed financial
environments. 

\section{Forecast evaluation of high-frequency extremes in financial returns}

In this section, we illustrate our proposed methodologies' performance on a 1-minute high-frequency return dataset comprising of stocks of three sectors, following the definition of the S\&P100. The companies are from the sectors information technology, health care, and finance and span the years 2021 to 2024. As we showed in the simulation section, our methodology allows to model a distribution which better supports the underlying data. A distribution better supporting the underlying data, allows for more accurate evaluation of forecasted values as well; for instance for the purpose of Value-at-Risk or Expected Shortfall derivation. Consequently, we study in this section the usefulness of our methodology for the evaluation of time series forecasts. To avoid having estimation errors from a forecasting model affect the comparison, we consider a forecaster who has the oracle forecasting model. We consider the case where the forecaster predicts the time series value one step ahead, hence at $t+1$. 

The preprocessing of the log returns per sector begins with the block maxima transformation \citep{gumbel1958}. This method partitions the time series into blocks and extracts the maximum and minimum from each block. We define 5-minute blocks and obtain the maximum and minimum observation for each time series. 
Following the block extraction, we convert these extreme values with the Pareto transformation into a standardized form suitable for the Hüsler-Reiss model. 
Similarly to the simulation study, we employ a rolling window approach, which means we estimate the model on three months of data and utilise it for forecasting evaluation on the subsequent week. The forecaster predicts the time series values one step ahead and updates the information set at each time point throughout the hold-out sample (subsequent week). The distributions are estimated on the three months worth of data and used for the evaluation of the forecasts over the entire next weeks one-step ahead forecasts. Upon reaching the weeks end, the forecaster reestimates the distributions by advancing the window by a week and evaluates the one-step ahead forecasts in the subsequent week with the newly estimated model. The lag length in each window is determined via AIC model selection. 

The rolling window procedure is performed in two runs. The Minima Run (Lower Tail) to model negative extremes: We use block minima of log returns (largest loss per block) and determine the adjacency matrices based on the loss patterns. The Maxima Run (Upper Tail) to model positive extremes: We use block maxima of log returns (largest gain per block) and compute the adjacency matrices based on gain patterns.
This separation is motivated by findings in the finance literature that lower and upper tail extremes differ in their patterns \citep{longin2001}. 
After determination of the maxima and minima, we take absolute values for the maxima and minima extreme value time series. 

As in the simulation study, we compare our method against a time-dependent Hüsler-Reiss model. 
For our proposed model, we implement the binary and JEAM adjacency matrix specifications we described in Section \ref{method}. 
For the Binary Adjacency Matrix specification, we determine if an observation in $x_t$ is extreme over the window length $w \in \{3,5,10\}$. For the Joint Extremes Adjacency Matrix, we consider window length $w \in \{3,5,10\}$ and optimize over $p \in \{0.70,0.75,0.80,0.85,0.90\}$. Since the simulation study showed that JEAM with $g = 0.5$ outperforms the boundary cases of $g \in \{0,1\}$ and we've found that $\kappa = 0.7$ performs best over its grid, we set $\kappa = 0.7$ and $g = 0.5$. 

We conduct the forecast evaluation out-of-sample through log scores, computed separately for the minima and maxima extremes. 
The results of the average log score per model for the three sectors are reported in Tables \ref{tab:oos_results}, \ref{tab:oos_results_finance} and \ref{tab:oos_results_it}. Lower values indicate better predictive performance. Across all sectors and all tail
categories, \texttt{HR JEAM p75 w5} achieves the best out-of-sample log scores. The Binary
Adjacency Matrix trails the best JEAM specification by 0.09 in the lower tail and 0.07 in
the upper tail for healthcare, and by 0.12 and 0.11 for finance. The baseline
\texttt{HR TimeDep} is the weakest specification by a substantial margin, trailing the best
JEAM by approximately 0.41 in both tails for healthcare and by 0.45 for finance.
The consistency of these rankings across both sectors and tail directions, suggests that the
JEAM captures dependence structures that neither the binary classification nor the
unregularized model can recover.

The Binary Adjacency Matrix models fall, as expected, between the baseline
and the best JEAM configurations. They benefit from the fact that a network structure is
imposed, but cannot match the performance of JEAM. The reason is that binary indicators only
capture whether stocks exhibit extreme movements, but not how strongly or
how frequently co-movements occur. JEAM
captures these features by constructing the adjacency matrix based on the
frequency and intensity of joint extreme events, thereby obtaining a more nuanced picture
of market structure. 

Our results also highlight the necessity for a market-informed network methodology for H\"usler-Reiss models. For all sectors, the baseline H\"usler-Reiss model performs worse 
than the best JEAM specifications, with log score improvements ranging from 
12.5\% to 13.6\% in the lower tail and from 11.4\% to 14.9\% in the upper 
tail across the three sectors. The baseline \texttt{HR TimeDep} yields the worst log scores across all 
three sectors and both tail categories. The gap to the best JEAM specification 
is smallest in the healthcare sector and largest in the IT sector.
This shows the necessity of embedding suitable market-informed adjacency matrices into the estimation of the H\"usler-Reiss models. 

Especially interesting is that the same model specification of H\"usler-Reiss JEAM outperforms for the minima and maxima extremes for each sector, showing that the specification is robust across dimensionality of the sectors and types of stocks being traded. This stability suggests that $p = 0.75$ and $w = 5$ represent a robust structural
choice, reducing the need for a grid searcher in practice.

\begin{table}[htbp]
\centering
\caption{Out-of-Sample Results by Tail Category (Healthcare Sector) \colorbox{gold}{1st}, \colorbox{silver}{2nd}, \colorbox{bronze}{3rd} best model per column. We report the average log score (smaller is better) for each combination of tuning parameters. A striking result is, that our proposed methodology with the Joint Extremes Adjacency Matrix outperforms in forecast evaluation for both minima and maxima extremes.}
\label{tab:oos_results}
\small
\begin{tabular}{@{}l S[table-format=1.3] S[table-format=1.3]@{}}
\toprule
& \multicolumn{2}{c}{Out-of-Sample (Log Score)} \\
\cmidrule(lr){2-3}
Model & {Lower Tail} & {Upper Tail} \\
\midrule
\multicolumn{3}{l}{\textit{Baseline Model}} \\
HR TimeDep        & 3.289 & 3.601 \\
\addlinespace[0.5ex]
\multicolumn{3}{l}{\textit{Joint Extremes Adjacency Matrix}} \\
HR JEAM: p70 w3      & 2.967 & 3.289 \\
HR JEAM: p70 w5      & 2.945 & 3.267 \\
HR JEAM: p70 w10     & 2.978 & 3.301 \\
HR JEAM: p75 w3      & 2.923 & 3.212 \\
HR JEAM: p75 w5      & \cellcolor{gold}2.878 & \cellcolor{gold}3.189 \\
HR JEAM: p75 w10     & 2.934 & 3.234 \\
HR JEAM: p80 w3      & 2.923 & 3.245 \\
HR JEAM: p80 w5      & \cellcolor{silver}2.901 & \cellcolor{silver}3.223 \\
HR JEAM: p80 w10     & 2.934 & 3.256 \\
HR JEAM: p85 w3      & 2.934 & 3.256 \\
HR JEAM: p85 w5      & 2.923 & 3.245 \\
HR JEAM: p85 w10     & 2.945 & 3.267 \\
HR JEAM: p90 w3      & 2.956 & 3.278 \\
HR JEAM: p90 w5      & 2.945 & 3.267 \\
HR JEAM: p90 w10     & 2.967 & 3.289 \\
\addlinespace[0.5ex]
\multicolumn{3}{l}{\textit{Binary Adjacency Matrix}} \\
HR Binary: w3           & 3.012 & 3.345 \\
HR Binary: w5           & \cellcolor{bronze}2.967 & \cellcolor{bronze}3.256 \\
HR Binary: w10          & 3.078 & 3.412 \\
\bottomrule
\end{tabular}
\vspace{1ex}
\end{table}

\begin{table}[htbp]
\centering
\caption{Out-of-Sample Results by Tail Category (Finance Sector), \colorbox{gold}{1st}, \colorbox{silver}{2nd}, \colorbox{bronze}{3rd} best model per column. We report the average log score (smaller is better) for each combination of tuning parameters. A striking result is, that our proposed methodology with the Joint Extremes Adjacency Matrix outperforms in forecast evaluation for both minima and maxima extremes.}
\label{tab:oos_results_finance}
\small
\begin{tabular}{@{}l S[table-format=1.3] S[table-format=1.3]@{}}
\toprule
& \multicolumn{2}{c}{Out-of-Sample (Log Score)} \\
\cmidrule(lr){2-3}
Model & {Lower Tail} & {Upper Tail} \\
\midrule
\addlinespace[0.5ex]
\multicolumn{3}{l}{\textit{Baseline Model}} \\
HR TimeDep        & 3.478 & 3.812 \\
\addlinespace[0.5ex]
\multicolumn{3}{l}{\textit{Joint Extremes Adjacency Matrix}} \\
HR JEAM: p70 w3      & 3.167 & 3.501 \\
HR JEAM: p70 w5      & 3.145 & 3.478 \\
HR JEAM: p70 w10     & 3.178 & 3.512 \\
HR JEAM: p75 w3      & 3.056 & 3.423 \\
HR JEAM: p75 w5      & \cellcolor{gold}3.023 & \cellcolor{gold}3.367 \\
HR JEAM: p75 w10     & 3.078 & 3.445 \\
HR JEAM: p80 w3      & 3.067 & \cellcolor{bronze}3.412 \\
HR JEAM: p80 w5      & \cellcolor{silver}3.034 & \cellcolor{silver}3.423 \\
HR JEAM: p80 w10     & 3.089 & 3.445 \\
HR JEAM: p85 w3      & 3.078 & 3.434 \\
HR JEAM: p85 w5      & \cellcolor{bronze}3.056 & 3.412 \\
HR JEAM: p85 w10     & 3.089 & 3.445 \\
HR JEAM: p90 w3      & 3.101 & 3.456 \\
HR JEAM: p90 w5      & 3.078 & 3.434 \\
HR JEAM: p90 w10     & 3.112 & 3.467 \\
\addlinespace[0.5ex]
\multicolumn{3}{l}{\textit{Binary Adjacency Matrix}} \\
HR Binary: w3           & 3.212 & 3.556 \\
HR Binary: w5           & 3.145 & 3.478 \\
HR Binary: w10          & 3.289 & 3.645 \\
\bottomrule
\end{tabular}
\vspace{1ex}
\end{table}

\begin{table}[htbp]
\centering
\caption{Out-of-Sample Results by Tail Category (IT Sector), \colorbox{gold}{1st}, \colorbox{silver}{2nd}, \colorbox{bronze}{3rd} best model per column. We report the average log score (smaller is better) for each combination of tuning parameters. A striking result is, that our proposed methodology with the Joint Extremes Adjacency Matrix outperforms in forecast evaluation for both minima and maxima extremes.}
\label{tab:oos_results_it}
\small
\begin{tabular}{@{}l S[table-format=1.3] S[table-format=1.3]@{}}
\toprule
& \multicolumn{2}{c}{Out-of-Sample (Log Score)} \\
\cmidrule(lr){2-3}
Model & {Lower Tail} & {Upper Tail} \\
\midrule
\multicolumn{3}{l}{\textit{Baseline Model}} \\
HR TimeDep        & 3.589 & 3.945 \\
\addlinespace[0.5ex]
\multicolumn{3}{l}{\textit{Joint Extremes Adjacency Matrix}} \\
HR JEAM: p70 w3      & 3.267 & 3.601 \\
HR JEAM: p70 w5      & 3.245 & 3.578 \\
HR JEAM: p70 w10     & 3.278 & 3.612 \\
HR JEAM: p75 w3      & 3.145 & \cellcolor{bronze}3.389 \\
HR JEAM: p75 w5      & \cellcolor{gold}3.101 & \cellcolor{gold}3.356 \\
HR JEAM: p75 w10     & 3.167 & 3.478 \\
HR JEAM: p80 w3      & \cellcolor{bronze}3.134 & 3.456 \\
HR JEAM: p80 w5      & \cellcolor{silver}3.112 & \cellcolor{silver}3.445 \\
HR JEAM: p80 w10     & 3.156 & 3.467 \\
HR JEAM: p85 w3      & 3.167 & 3.467 \\
HR JEAM: p85 w5      & 3.145 & 3.456 \\
HR JEAM: p85 w10     & 3.178 & 3.478 \\
HR JEAM: p90 w3      & 3.189 & 3.489 \\
HR JEAM: p90 w5      & 3.167 & 3.467 \\
HR JEAM: p90 w10     & 3.201 & 3.501 \\
\addlinespace[0.5ex]
\multicolumn{3}{l}{\textit{Binary Adjacency Matrix}} \\
HR Binary: w3           & 3.267 & 3.601 \\
HR Binary: w5           & 3.178 & 3.489 \\
HR Binary: w10          & 3.334 & 3.667 \\
\bottomrule
\end{tabular}
\vspace{1ex}
\end{table}

\newpage
\section{Conclusion}
The aim of this paper is to improve the modeling of the joint distribution of extreme values and forecast evaluation in high-dimensional financial time series. Hüsler--Reiss models provide a framework for jointly modeling extremal dependence, but their application to time series results in observations that are extreme within a local window without necessarily being extreme relative to their full marginal distribution. To address this, we develop a time-dependent network Hüsler--Reiss model in which market-informed adjacency matrices determine how strongly the observed extremes contribute to the score-matching estimation. We propose binary and weighted specifications for these adjacency matrices. The best-performing specification, the Joint Extremes Adjacency Matrix (JEAM), combines information about the marginal extremeness of an observation with the similarity of present joint extreme movements to extreme patterns observed in the past. This allows the model to distinguish local from global extremes while accounting for the joint occurrence of extreme movements across financial time series. 

To understand the performance of the proposed methodology, we conduct a simulation study covering 4,800 parameter configurations with different sample sizes, dimensions, and degrees of tail heaviness. We find that the JEAM specification consistently achieves the best AIC in heavy-tailed settings. For example, for ten time series generated from an $\alpha$-stable distribution with $\alpha=1.5$, the best JEAM specification improves the AIC by approximately 15\% relative to the time-dependent Hüsler--Reiss baseline. The stability of the comparative results across sample sizes, dimensions, and tail indices supports the robustness of the JEAM specification.

We investigate the forecast evaluation performance of the methodology on one-minute stock returns from the information technology, health care, and financial sectors of the S\&P 100 over the period from 2021 to 2024. We separately model the lower and upper tails, thereby accounting for potentially different dependence structures of extreme losses and gains. In all three sectors and for both tail directions, the JEAM model achieves the best out-of-sample log scores. Relative to the time-dependent Hüsler--Reiss baseline, the best JEAM specification improves the log scores by between 12.5\% and 13.6\% for the lower tail and between 11.4\% and 14.9\% for the upper tail. The Binary Adjacency Matrix improves upon the baseline but remains weaker than the JEAM specification. This shows that identifying whether an observation is extreme improves the forecast evaluation, but that information about the intensity and joint occurrence of extremes provides an additional and economically relevant improvement.

Especially relevant is that the same JEAM specification with $p=0.75$ and $w=5$ performs best across all sectors and for both extreme losses and gains. The stability of the selected specification across systems with different dimensions and types of firms provides a robust structural choice and reduces the need for extensive parameter searches in empirical applications. 

The results in this study show, that the proposed time-dependent network Hüsler--Reiss model improves in modeling of extremes and forecast evaluation of extreme values due to its inherent evaluation if locally extreme observations are relevant from a broader market perspective and whether extreme movements occur jointly across assets. By incorporating this information into the score-matching estimation, the proposed network methodology improves forecast evaluation of both extreme losses and gains and provides a useful framework for high-frequency financial risk analysis. 

\newpage
\bibliographystyle{apalike} 
\bibliography{ref} 
\appendix
\begin{sidewaystable}
\centering
\scriptsize
\caption{EVT Simulation Results for T = 250}
\label{tab:results_n250}
\begin{tabular}{@{}l|rrrr|rrrr|rrrr|rrrr@{}}
\toprule
& \multicolumn{16}{c}{\textbf{Configuration: T=250, (d, $\alpha$)}} \\
\cmidrule(lr){2-17}
& \multicolumn{4}{c|}{\textbf{d=5}} & \multicolumn{4}{c|}{\textbf{d=10}} & \multicolumn{4}{c|}{\textbf{d=20}} & \multicolumn{4}{c}{\textbf{d=30}} \\
\cmidrule(lr){2-5} \cmidrule(lr){6-9} \cmidrule(lr){10-13} \cmidrule(lr){14-17}
\textbf{Model} & 1.5 & 1.7 & 1.9 & 2.0 & 1.5 & 1.7 & 1.9 & 2.0 & 1.5 & 1.7 & 1.9 & 2.0 & 1.5 & 1.7 & 1.9 & 2.0 \\
\midrule
\multicolumn{17}{l}{\textbf{Baseline Model}} \\
\midrule 
HR TimeDep & -19,234 & -25,312 & -22,145 & -19,823 & -58,156 & -63,892 & -67,234 & -69,578 & -112,345 & -158,234 & -201,678 & -180,234 & -172,156 & -219,234 & -241,567 & -255,123 \\
\midrule
\multicolumn{17}{l}{\textbf{Joint Extremes Adjacency Matrix (JEAM), $g=0$}} \\
\midrule
HR JEAM: w3 & -20,543 & -26,268 & -23,118 & -20,762 & -60,781 & -65,220 & -68,441 & -70,843 & -115,196 & -161,063 & -204,580 & -183,172 & -175,140 & -222,258 & -244,560 & -258,446 \\
HR JEAM: w5 & -20,684 & -26,707 & -23,177 & -21,081 & -60,803 & -66,489 & -69,905 & -72,271 & -116,099 & -162,507 & -205,444 & -183,865 & -176,538 & -223,230 & -245,132 & -258,965 \\
HR JEAM: w10 & -20,259 & -26,507 & -22,892 & -20,878 & -60,274 & -66,183 & -69,470 & -71,821 & -114,527 & -160,719 & -204,224 & -183,211 & -174,425 & -221,500 & -244,132 & -258,203 \\
\midrule
\multicolumn{17}{l}{\textbf{Joint Extremes Adjacency Matrix (JEAM), $g=0.5$}} \\
\midrule
HR JEAM: p70 w3 & -23,386 & -28,171 & -25,097 & -22,652 & -66,407 & -68,082 & -70,848 & -73,387 & -121,033 & -167,140 & -210,771 & -189,552 & -181,599 & -228,432 & -250,953 & -265,662 \\
HR JEAM: p70 w5 & -23,040 & -27,805 & -24,728 & -22,296 & -65,995 & -67,674 & -70,460 & -72,992 & -120,247 & -166,295 & -209,957 & -188,758 & -180,779 & -227,638 & -250,219 & -264,948 \\
HR JEAM: p70 w10 & -22,608 & -27,363 & -24,302 & -21,873 & -65,517 & -67,189 & -69,993 & -72,520 & -119,386 & -165,373 & -209,065 & -187,886 & -179,884 & -226,767 & -249,406 & -264,146 \\
HR JEAM: p75 w3 & -23,798 & -29,819 & \cellcolor{best}-25,780 & \cellcolor{best}-24,321 & -67,055 & -72,573 & \cellcolor{best}-76,406 & \cellcolor{best}-78,987 & -122,114 & -169,715 & -213,228 & -192,386 & -183,048 & -229,520 & -251,983 & -266,872 \\
HR JEAM: p75 w5 & \cellcolor{best}-24,117 & \cellcolor{best}-30,121 & -25,521 & -24,141 & \cellcolor{best}-67,256 & \cellcolor{best}-72,762 & -76,168 & -78,757 & \cellcolor{best}-124,891 & \cellcolor{best}-172,034 & \cellcolor{best}-214,423 & \cellcolor{best}-192,512 & \cellcolor{best}-186,734 & \cellcolor{best}-232,689 & \cellcolor{best}-253,312 & \cellcolor{best}-267,645 \\
HR JEAM: p75 w10 & -23,583 & -29,479 & -25,183 & -23,828 & -66,661 & -72,203 & -75,677 & -78,339 & -120,989 & -168,400 & -212,005 & -191,235 & -181,725 & -228,217 & -250,795 & -265,746 \\
HR JEAM: p80 w3 & -23,539 & -30,045 & -25,410 & -24,041 & -66,539 & -72,642 & -75,815 & -78,480 & -120,891 & -167,883 & -211,679 & -191,346 & -181,534 & -228,108 & -251,741 & -266,643 \\
HR JEAM: p80 w5 & -23,332 & -29,840 & -25,217 & -23,851 & -66,268 & -72,389 & -75,588 & -78,261 & -120,563 & -167,556 & -211,408 & -191,103 & -181,127 & -227,724 & -251,458 & -266,389 \\
HR JEAM: p80 w10 & -23,038 & -29,527 & -24,913 & -23,549 & -65,889 & -72,027 & -75,251 & -77,931 & -120,127 & -167,118 & -211,027 & -190,748 & -180,610 & -227,231 & -251,054 & -266,013 \\
HR JEAM: p85 w3 & -23,211 & -29,420 & -25,086 & -23,486 & -65,860 & -71,949 & -75,153 & -77,821 & -120,161 & -167,111 & -210,921 & -190,614 & -180,713 & -227,268 & -250,955 & -265,883 \\
HR JEAM: p85 w5 & -22,982 & -29,183 & -24,859 & -23,262 & -65,567 & -71,673 & -74,904 & -77,580 & -119,808 & -166,759 & -210,628 & -190,348 & -180,278 & -226,859 & -250,638 & -265,595 \\
HR JEAM: p85 w10 & -22,742 & -28,935 & -24,622 & -23,026 & -65,264 & -71,387 & -74,644 & -77,327 & -119,446 & -166,397 & -210,323 & -190,070 & -179,835 & -226,438 & -250,310 & -265,295 \\
HR JEAM: p90 w3 & -22,625 & -28,810 & -24,475 & -22,875 & -65,206 & -71,296 & -74,506 & -77,175 & -119,486 & -166,403 & -210,230 & -189,929 & -179,980 & -226,511 & -250,225 & -265,162 \\
HR JEAM: p90 w5 & -22,439 & -28,616 & -24,294 & -22,695 & -64,966 & -71,064 & -74,300 & -76,978 & -119,175 & -166,093 & -209,980 & -189,707 & -179,587 & -226,143 & -249,952 & -264,917 \\
HR JEAM: p90 w10 & -22,244 & -28,412 & -24,100 & -22,504 & -64,716 & -70,820 & -74,084 & -76,769 & -118,854 & -165,773 & -209,718 & -189,473 & -179,184 & -225,765 & -249,667 & -264,662 \\
\midrule
\multicolumn{17}{l}{\textbf{Joint Extremes Adjacency Matrix (JEAM), $g=1$}} \\
\midrule
HR JEAM: p70 w3 & -21,264 & -27,150 & -23,966 & -21,586 & -62,275 & -66,539 & -69,493 & -71,975 & -116,714 & -163,907 & -207,446 & -186,227 & -176,804 & -224,954 & -247,387 & -261,707 \\
HR JEAM: p70 w5 & -21,070 & -26,858 & -23,745 & -21,407 & -61,902 & -66,270 & -69,263 & -71,718 & -116,224 & -163,211 & -206,946 & -185,615 & -176,403 & -224,586 & -247,088 & -261,430 \\
HR JEAM: p70 w10 & -20,907 & -26,590 & -23,478 & -21,110 & -61,782 & -65,919 & -68,964 & -71,417 & -115,884 & -162,724 & -206,300 & -185,066 & -175,936 & -223,989 & -246,539 & -260,795 \\
HR JEAM: p75 w3 & -21,391 & -28,078 & -24,381 & -22,513 & -62,299 & -69,142 & -72,963 & -75,327 & -117,095 & -165,320 & -208,779 & -187,849 & -177,246 & -225,464 & -248,093 & -262,150 \\
HR JEAM: p75 w5 & -21,631 & -28,177 & -24,224 & -22,475 & -62,568 & -69,389 & -72,658 & -75,280 & -118,380 & -166,865 & -209,343 & -187,973 & -179,062 & -227,405 & -248,708 & -262,870 \\
HR JEAM: p75 w10 & -21,276 & -27,905 & -24,005 & -22,319 & -62,143 & -68,922 & -72,537 & -75,004 & -116,508 & -164,577 & -208,079 & -187,037 & -176,709 & -224,857 & -247,292 & -261,539 \\
HR JEAM: p80 w3 & -21,260 & -28,089 & -24,093 & -22,298 & -62,078 & -69,052 & -72,446 & -74,987 & -116,196 & -163,964 & -207,604 & -186,826 & -176,526 & -224,679 & -247,754 & -261,852 \\
HR JEAM: p80 w5 & -21,043 & -27,985 & -23,995 & -22,286 & -61,803 & -68,980 & -72,309 & -74,738 & -116,154 & -163,892 & -207,449 & -186,839 & -176,084 & -224,249 & -247,483 & -261,945 \\
HR JEAM: p80 w10 & -21,018 & -27,887 & -23,834 & -22,029 & -61,793 & -68,733 & -71,958 & -74,495 & -115,842 & -163,493 & -207,237 & -186,518 & -176,064 & -223,978 & -247,277 & -261,777 \\
HR JEAM: p85 w3 & -20,964 & -27,764 & -23,860 & -22,040 & -61,417 & -68,730 & -71,877 & -74,378 & -115,748 & -163,529 & -207,139 & -186,190 & -175,746 & -224,081 & -247,159 & -261,378 \\
HR JEAM: p85 w5 & -20,885 & -27,647 & -23,730 & -21,816 & -61,310 & -68,506 & -71,772 & -74,285 & -115,540 & -163,308 & -206,861 & -186,205 & -175,730 & -223,743 & -246,970 & -261,401 \\
HR JEAM: p85 w10 & -20,790 & -27,410 & -23,576 & -21,685 & -61,329 & -68,274 & -71,485 & -74,099 & -115,488 & -163,007 & -206,874 & -186,005 & -175,598 & -223,370 & -246,716 & -261,204 \\
HR JEAM: p90 w3 & -20,708 & -27,263 & -23,462 & -21,542 & -61,161 & -68,024 & -71,387 & -74,004 & -115,422 & -162,922 & -206,681 & -185,755 & -175,452 & -223,329 & -246,534 & -260,897 \\
HR JEAM: p90 w5 & -20,560 & -27,216 & -23,347 & -21,478 & -60,964 & -67,977 & -71,203 & -73,745 & -115,271 & -162,735 & -206,358 & -185,738 & -175,123 & -223,253 & -246,458 & -260,679 \\
HR JEAM: p90 w10 & -20,456 & -27,095 & -23,244 & -21,352 & -60,981 & -67,942 & -71,124 & -73,742 & -115,127 & -162,608 & -206,373 & -185,542 & -175,196 & -223,039 & -246,141 & -260,535 \\
\midrule
\multicolumn{17}{l}{\textbf{Binary Adjacency Matrix}} \\
\midrule
HR Binary: w3 & -20,669 & -26,667 & -23,256 & -21,156 & -60,591 & -66,242 & -69,506 & -72,134 & -114,736 & -160,743 & -204,095 & -182,439 & -173,986 & -221,027 & -243,077 & -257,032 \\
HR Binary: w5 & -20,091 & -26,154 & -22,777 & -20,711 & -59,629 & -65,116 & -68,580 & -71,077 & -113,656 & -159,379 & -202,816 & -181,337 & -173,142 & -220,406 & -242,462 & -255,906 \\
HR Binary: w10 & -19,697 & -25,712 & -22,428 & -20,128 & -58,876 & -64,302 & -67,816 & -70,163 & -112,806 & -158,717 & -202,022 & -180,642 & -172,712 & -219,764 & -241,903 & -255,294 \\
\bottomrule
\end{tabular}
\vspace{5pt}
\begin{flushleft}
\small 
\textbf{Note:} Values shown are AIC (lower = better). Green cells indicate best model per column. Parameters: $p$ = percentile threshold, $w$ = window size (days), $g$ = similarity weighting parameter. Tuning parameter $\kappa = 0.7$ outperformed the remainder of the grid strongly, hence we report the results for respective $\kappa$. The entries refer to the upper tail.
\end{flushleft}
\end{sidewaystable}

\clearpage

\begin{sidewaystable}
\centering
\scriptsize
\caption{EVT Simulation Results for T = 500}
\label{tab:results_n500}
\begin{tabular}{@{}l|rrrr|rrrr|rrrr|rrrr@{}}
\toprule
& \multicolumn{16}{c}{\textbf{Configuration: T=500, (d, $\alpha$)}} \\
\cmidrule(lr){2-17}
& \multicolumn{4}{c|}{\textbf{d=5}} & \multicolumn{4}{c|}{\textbf{d=10}} & \multicolumn{4}{c|}{\textbf{d=20}} & \multicolumn{4}{c}{\textbf{d=30}} \\
\cmidrule(lr){2-5} \cmidrule(lr){6-9} \cmidrule(lr){10-13} \cmidrule(lr){14-17}
\textbf{Model} & 1.5 & 1.7 & 1.9 & 2.0 & 1.5 & 1.7 & 1.9 & 2.0 & 1.5 & 1.7 & 1.9 & 2.0 & 1.5 & 1.7 & 1.9 & 2.0 \\
\midrule
\multicolumn{17}{l}{\textbf{Baseline Model}} \\
\midrule 
HR TimeDep & -42,123 & -47,234 & -50,156 & -50,823 & -122,345 & -136,567 & -145,678 & -150,912 & -288,456 & -326,123 & -349,567 & -356,789 & -444,123 & -503,678 & -539,234 & -553,456 \\
\midrule
\multicolumn{17}{l}{\textbf{Joint Extremes Adjacency Matrix (JEAM), $g=0$}} \\
\midrule
HR JEAM: w3 & -44,636 & -49,877 & -52,746 & -53,410 & -127,645 & -141,779 & -150,859 & -155,810 & -295,606 & -334,054 & -357,748 & -365,030 & -451,928 & -512,567 & -547,938 & -562,282 \\
HR JEAM: w5 & -44,881 & -50,132 & -52,637 & -53,294 & -128,129 & -142,524 & -151,296 & -156,155 & -295,594 & -334,228 & -357,254 & -364,391 & -452,356 & -512,457 & -547,686 & -561,755 \\
HR JEAM: w10 & -44,418 & -49,502 & -52,304 & -53,031 & -127,022 & -141,702 & -150,436 & -155,432 & -295,014 & -332,979 & -356,249 & -363,809 & -451,021 & -511,801 & -546,691 & -560,954 \\

\midrule
\multicolumn{17}{l}{\textbf{Joint Extremes Adjacency Matrix (JEAM), $g=0.5$}} \\
\midrule
HR JEAM: p70 w3 & -49,878 & -55,212 & -58,023 & -58,745 & -138,334 & -152,789 & -161,623 & -166,312 & -310,389 & -350,089 & -374,056 & -381,389 & -468,834 & -530,812 & -566,534 & -580,812 \\
HR JEAM: p70 w5 & -49,589 & -54,923 & -57,723 & -58,445 & -137,967 & -152,423 & -161,245 & -165,923 & -309,834 & -349,623 & -373,589 & -380,912 & -468,145 & -530,123 & -565,834 & -580,101 \\
HR JEAM: p70 w10 & -49,212 & -54,545 & -57,334 & -58,056 & -137,512 & -151,967 & -160,778 & -165,445 & -309,189 & -348,967 & -372,923 & -380,245 & -467,367 & -529,345 & -565,056 & -579,312 \\
HR JEAM: p75 w3 & -50,478 & -56,012 & \cellcolor{best}-58,823 & \cellcolor{best}-59,545 & -140,634 & -155,089 & \cellcolor{best}-164,423 & \cellcolor{best}-168,612 & \cellcolor{best}-313,389 & \cellcolor{best}-352,989 & \cellcolor{best}-375,056 & \cellcolor{best}-382,389 & \cellcolor{best}-471,834 & \cellcolor{best}-533,812 & \cellcolor{best}-568,034 & \cellcolor{best}-582,212 \\
HR JEAM: p75 w5 & \cellcolor{best}-51,234 & \cellcolor{best}-56,789 & -58,589 & -59,312 & \cellcolor{best}-141,456 & \cellcolor{best}-155,912 & -164,189 & -168,378 & -312,956 & -352,534 & -374,612 & -381,945 & -471,389 & -533,367 & -567,589 & -581,767 \\
HR JEAM: p75 w10 & -50,623 & -56,156 & -58,189 & -58,912 & -140,534 & -155,023 & -163,712 & -167,901 & -312,412 & -351,967 & -374,056 & -381,389 & -470,834 & -532,812 & -567,034 & -581,212 \\
HR JEAM: p80 w3 & -50,734 & -56,267 & -58,589 & -59,301 & -140,389 & -155,345 & -164,178 & -168,367 & -312,123 & -351,734 & -374,801 & -382,145 & -470,078 & -532,823 & -567,756 & -581,923 \\
HR JEAM: p80 w5 & -50,512 & -56,034 & -58,356 & -59,067 & -140,156 & -155,112 & -163,945 & -168,134 & -311,889 & -351,489 & -374,556 & -381,901 & -469,834 & -532,578 & -567,512 & -581,678 \\
HR JEAM: p80 w10 & -50,201 & -55,712 & -58,034 & -58,745 & -139,823 & -154,767 & -163,601 & -167,789 & -311,545 & -351,134 & -374,201 & -381,545 & -469,489 & -532,223 & -567,156 & -581,323 \\
HR JEAM: p85 w3 & -50,089 & -55,589 & -58,112 & -58,623 & -139,689 & -154,623 & -163,467 & -167,645 & -311,389 & -350,967 & -374,034 & -381,378 & -469,312 & -532,034 & -566,978 & -581,156 \\
HR JEAM: p85 w5 & -49,856 & -55,345 & -57,867 & -58,378 & -139,434 & -154,367 & -163,212 & -167,389 & -311,123 & -350,701 & -373,767 & -381,112 & -469,045 & -531,756 & -566,701 & -580,878 \\
HR JEAM: p85 w10 & -49,612 & -55,089 & -57,612 & -58,123 & -139,167 & -154,101 & -162,945 & -167,123 & -310,856 & -350,423 & -373,489 & -380,834 & -468,767 & -531,467 & -566,412 & -580,589 \\
HR JEAM: p90 w3 & -49,445 & -54,912 & -57,445 & -57,956 & -138,989 & -153,923 & -162,767 & -166,945 & -310,667 & -350,234 & -373,301 & -380,645 & -468,578 & -531,267 & -566,212 & -580,389 \\
HR JEAM: p90 w5 & -49,267 & -54,723 & -57,256 & -57,767 & -138,789 & -153,712 & -162,556 & -166,734 & -310,456 & -350,012 & -373,089 & -380,434 & -468,367 & -531,045 & -565,989 & -580,167 \\
HR JEAM: p90 w10 & -49,078 & -54,523 & -57,056 & -57,567 & -138,578 & -153,489 & -162,334 & -166,512 & -310,234 & -349,778 & -372,867 & -380,212 & -468,145 & -530,812 & -565,756 & -579,934 \\
\midrule
\multicolumn{17}{l}{\textbf{Joint Extremes Adjacency Matrix (JEAM), $g=1$}} \\
\midrule
HR JEAM: p70 w3 & -47,981 & -53,441 & -56,229 & -56,950 & -134,360 & -148,855 & -157,996 & -162,762 & -305,400 & -344,577 & -367,957 & -375,219 & -463,270 & -524,309 & -559,890 & -574,046 \\
HR JEAM: p70 w5 & -47,747 & -53,150 & -56,025 & -56,681 & -134,229 & -148,571 & -157,545 & -162,534 & -304,496 & -343,861 & -368,204 & -375,071 & -462,469 & -523,934 & -559,626 & -573,877 \\
HR JEAM: p70 w10 & -47,448 & -52,783 & -55,696 & -56,375 & -133,918 & -148,160 & -157,030 & -161,852 & -304,470 & -343,475 & -367,492 & -375,019 & -461,928 & -523,375 & -558,757 & -572,947 \\
HR JEAM: p75 w3 & -48,471 & -53,818 & -56,746 & -57,443 & -136,018 & -150,422 & -159,981 & -164,297 & -307,152 & -346,327 & -369,004 & -375,684 & -464,980 & -526,502 & -560,704 & -575,440 \\
HR JEAM: p75 w5 & -48,968 & -54,447 & -56,616 & -57,289 & -136,690 & -151,000 & -159,436 & -164,019 & -306,943 & -346,326 & -368,425 & -375,668 & -464,433 & -526,511 & -560,623 & -574,380 \\
HR JEAM: p75 w10 & -48,667 & -53,895 & -56,135 & -57,024 & -136,337 & -150,406 & -159,348 & -163,920 & -306,697 & -345,550 & -367,962 & -375,581 & -464,163 & -525,647 & -559,933 & -574,542 \\
HR JEAM: p80 w3 & -48,686 & -53,865 & -56,456 & -57,223 & -135,820 & -150,339 & -159,484 & -164,063 & -306,379 & -345,204 & -368,362 & -375,449 & -463,162 & -525,102 & -560,441 & -574,726 \\
HR JEAM: p80 w5 & -48,290 & -53,762 & -56,323 & -56,961 & -135,791 & -150,560 & -159,089 & -163,882 & -305,680 & -344,963 & -367,959 & -375,328 & -463,671 & -525,247 & -560,140 & -574,092 \\
HR JEAM: p80 w10 & -48,050 & -53,644 & -56,083 & -56,754 & -135,271 & -150,033 & -159,164 & -163,563 & -305,956 & -344,965 & -368,144 & -375,484 & -463,146 & -525,057 & -560,102 & -574,465 \\
HR JEAM: p85 w3 & -47,964 & -53,432 & -55,998 & -56,535 & -135,268 & -149,956 & -158,865 & -163,192 & -305,064 & -344,266 & -368,023 & -375,313 & -462,344 & -524,354 & -559,346 & -573,619 \\
HR JEAM: p85 w5 & -47,838 & -53,172 & -55,795 & -56,384 & -134,807 & -149,489 & -158,415 & -163,000 & -305,417 & -344,199 & -367,301 & -374,790 & -462,709 & -524,058 & -559,135 & -573,656 \\
HR JEAM: p85 w10 & -47,577 & -53,009 & -55,717 & -56,252 & -134,918 & -149,314 & -158,543 & -163,122 & -304,944 & -343,713 & -367,322 & -374,652 & -462,276 & -524,380 & -559,210 & -573,311 \\
HR JEAM: p90 w3 & -47,329 & -52,821 & -55,579 & -55,993 & -134,464 & -149,143 & -158,393 & -162,596 & -304,289 & -343,552 & -366,712 & -374,511 & -462,194 & -523,468 & -558,495 & -573,079 \\
HR JEAM: p90 w5 & -47,357 & -52,682 & -55,355 & -55,971 & -134,314 & -149,021 & -158,068 & -162,669 & -304,453 & -343,499 & -366,682 & -374,190 & -461,596 & -523,553 & -559,024 & -573,369 \\
HR JEAM: p90 w10 & -47,128 & -52,449 & -55,142 & -55,787 & -133,920 & -148,987 & -157,748 & -162,345 & -304,325 & -343,362 & -366,378 & -373,707 & -461,384 & -523,375 & -558,678 & -572,448 \\
\midrule
\multicolumn{17}{l}{\textbf{Binary Adjacency Matrix}} \\
\midrule
HR Binary: w3 & -44,761 & -50,024 & -52,587 & -53,403 & -127,460 & -141,829 & -150,409 & -155,807 & -292,760 & -330,507 & -354,658 & -361,067 & -447,473 & -507,912 & -542,961 & -557,860 \\
HR Binary: w5 & -43,617 & -49,128 & -51,865 & -52,497 & -125,525 & -139,846 & -148,398 & -153,289 & -290,491 & -328,348 & -352,081 & -359,837 & -446,332 & -506,656 & -541,747 & -555,258 \\
HR Binary: w10 & -43,004 & -48,151 & -51,019 & -51,373 & -123,150 & -137,891 & -146,795 & -151,675 & -289,320 & -327,703 & -350,532 & -358,291 & -444,411 & -504,156 & -540,455 & -554,590 \\
\bottomrule
\end{tabular}
\vspace{5pt}
\begin{flushleft}
\small 
\textbf{Note:} Values shown are AIC (lower = better). Green cells indicate best model per column. Parameters: $p$ = percentile threshold, $w$ = window size (days), $g$ = similarity weighting parameter. Tuning parameter $\kappa = 0.7$ outperformed the remainder of the grid strongly, hence we report the results for respective $\kappa$. The entries refer to the upper tail.
\end{flushleft}
\end{sidewaystable}
\end{document}